\documentclass[prl,twocolumn,superscriptaddress,showpacs]{revtex4}
\usepackage{graphicx}
\usepackage{epsfig}           
\usepackage{bm}                 
\usepackage{amsmath}            
\usepackage{amsfonts}
\usepackage{amssymb}
\usepackage{latexsym}           
\usepackage{natbib}

\begin{document}

\title{Single-qubit optical quantum fingerprinting}

\author{Rolf T. Horn}
\affiliation{Institute for Quantum Information Science, University of Calgary, Alberta T2N 1N4, Canada}
\author{S. A. Babichev}
\affiliation{Institute for Quantum Information Science, University of Calgary, Alberta T2N 1N4, Canada}
\affiliation{Fachbereich Physik, Universit\"at Konstanz, D-78457 Konstanz, Germany}
\author{Karl-Peter Marzlin}
\affiliation{Institute for Quantum Information Science, University of Calgary, Alberta T2N 1N4, Canada}
\author{A. I. Lvovsky}
\affiliation{Institute for Quantum Information Science, University of Calgary, Alberta T2N 1N4, Canada}
\affiliation{Fachbereich Physik, Universit\"at Konstanz, D-78457 Konstanz, Germany}
\author{Barry C.\ Sanders}
\affiliation{Institute for Quantum Information Science, University of Calgary, Alberta T2N 1N4, Canada}

\date{\today}

\begin{abstract}

We analyze and demonstrate the feasibility and superiority of linear optical single-qubit fingerprinting
over its classical counterpart. For  one-qubit fingerprinting of two-bit messages, 
we prepare `tetrahedral' qubit states experimentally and show that they meet the requirements for
quantum fingerprinting to exceed the classical capability. We
prove that shared entanglement permits 100\%
reliable quantum fingerprinting, which will outperform classical
fingerprinting even with arbitrary amounts of shared randomness.

\end{abstract}

\pacs{03.67.Hk, 42.50.Dv}

\maketitle

\emph{Introduction.--} 
Quantum communication can significantly improve on the resource requirements 
compared to classical communication~\cite{Bra03}.
Fingerprinting,
which enables an efficient way of inferring whether longer messages are 
identical or not, is a particularly striking example as quantum fingerprinting 
offers an exponential reduction
of resources compared to classical fingerprinting~\cite{Buh01}.
In fact, even for single-qubit fingerprinting one can demonstrate an
advantage of quantum protocols with respect to classical ones~\cite{Bea04}.
Here we establish the feasibility of single-qubit optical quantum fingerprinting,
by theoretical analysis and also by experimentally generating and 
assessing the appropriate quantum optical states for encoding.
In particular we (i)~develop an optical protocol for single-qubit fingerprinting,
(ii)~show that two-photon coincidence measurements suffice as the experimental
test for comparing fingerprints, (iii)~prove that one
shared entangled bit between Alice and Bob allows zero-error
quantum fingerprinting which outperforms classical fingerprinting even
with unlimited shared randomness between Alice and Bob,
and (iv)~present experimental results on the supply of fingerprint 
states that demonstrates the
feasibility of the protocol.
Our results open the prospect of experimental quantum communication complexity;
although here we focus on single-qubit fingerprinting and correlated photon pairs,
scalability will become possible as multiphoton entanglement 
capabilities improve~\cite{Eib03}.

Within the simultaneous message passing model~\cite{Yao79},
fingerprinting is constructed as follows. Two parties, Alice (A) and Bob
(B), receive classical $n$-bit message inputs $x$ and $y$ from a supplier
Sapna (S). Alice and Bob wish to test their messages for equality
but are forbidden to communicate with each other.  They can
however communicate with a third party Roger (R).  Communication
is expensive, so Alice and Bob create (classical or quantum) fingerprints of
length~$g$ for their respective messages, which they
send to Roger. Roger's goal is to generate a single bit value $z$
which provides the best inference of the function
\begin{equation}
\label{Eq} \text{EQ}(x,y)=\left\{ \begin{array}{ll}0&\text{if
}x\neq y\\ 1&\text{if }x=y
    \end{array}\right.,
\end{equation}
and Roger is successful if $z=\text{EQ}(x,y)$. Each message belongs to a set
$M=\{0,\ldots,m-1\}$ comprised of $m$ different messages
represented as bit strings of length 
$n\equiv\lceil{\text{log}_2{m}}\rceil$) and each
fingerprint to a set $F=\{0,\ldots,f-1\}$ of $f$ different fingerprints.
Classically, $g=\lceil{\text{log}_2{f}}\rceil$ and $F=\{0,1\}^{g}$ while in the
quantum case $F\subset\mathcal{H}_2^{g}$ for
$\mathcal{H}_2=\text{span}\{|0\rangle,|1\rangle\}$. 
The protocol is evaluated according to the worst
case scenario (WCS), in which Sapna always sends message pairs
for which the probability for $z\neq\text{EQ}(x,y)$ is maximized
(i.e. performance
in the WCS corresponds to the `guarantee' on the protocol).

We consider two experimental scenarios: 1)~Alice and Bob each
simultaneously receive unentangled single photons~\cite{NJP04}
with polarization states expressed in the logical basis
$|0\rangle$ and $|1\rangle$ and, 2)~Alice and Bob share a source
of entangled photon pairs in the singlet Bell state
$|\Psi^-\rangle  \equiv (|0,1 \rangle -|1, 0\rangle )/\sqrt{2}$. 
In the first scenario we are able to show that a linear
optical single-qubit quantum fingerprinting protocol 
outperforms classical fingerprinting without a shared resource. In
the second scenario, Alice and Bob share entanglement, and we show
that this protocol can yield perfect one-qubit fingerprinting for
$m=4$, outperforming one-bit fingerprinting for $m=4$ with an
arbitrary amount of shared randomness.

\emph{Encoding:--} For any message $w\in M$ that Alice or Bob
receive, they transform their qubit to a unique $|\Omega_w\rangle$
with
$|\Omega\equiv(\theta,\phi)\rangle\equiv\cos\tfrac{\theta}{2}|0\rangle
 + \exp(\text{i}\phi)\sin\tfrac{\theta}{2}|1\rangle$.
The state can be understood geometrically by identifying $\theta$
and $\phi$ with azimuthal and polar angles of the (Bloch) sphere.
We assume that Alice and Bob employ the same mapping:
$x=y\Leftrightarrow|\Omega_{x}\rangle=|\Omega_{y}\rangle$. Quantum
fingerprinting allows $m$ different qubit states so each message
is distinctly encoded, but the distinguishability of these
distinct states diminishes as $m$ increases, with
indistinguishability quantified by
$\delta(\Omega',\Omega)\equiv|\langle\Omega'|\Omega\rangle|^2
    = |\cos\tfrac{\theta'}{2}\cos\tfrac{\theta}{2} +
    \exp[\text{i}(\phi-\phi')]\sin\tfrac{\theta'}{2}
    \sin\tfrac{\theta}{2}|^2$. 
Because of a nonzero overlap, Roger can misinterpret
two different messages as identical. In the WCS, 
the corresponding
error rate depends on
$\delta_{\text{max}}\equiv\max_{(w \neq w')}
\delta(\Omega_{w'},\Omega_w)$, and the strategy for qubit encoding
should minimize $\delta_{\text{max}}$.

Single-qubit fingerprinting is especially interesting because of
its current feasibility. To demonstrate this, we analyze the case
$m=4$ ($n=2$). In this case $\delta_{\text{max}}$ is minimized by the 
following set of four states,
\begin{align}
\label{M=4protocol}
 F=\{|&\Omega_w\rangle; \Omega_0\equiv(\theta_0,\phi_0)=(0,0)\text{ or }\nonumber \\
&\Omega_w=(2\cos^{-1}\frac{1}{\sqrt{3}},\frac{2\pi}{3} w)\text{
for }w=1,2,3\},
\end{align}
and $\delta=\tfrac{1}{3}$ for all pairs of different states
\cite{Cla01,Dav78}. We refer to the states~(\ref{M=4protocol}) as
`tetrahedral states' because the four states form the vertices of
a tetrahedron on the Bloch sphere \cite{Dav78}.

\emph{Protocol.--} Alice and Bob map their two-bit messages to the
tetrahedral states, and Roger's task is to assess EQ$(x,y)$ by
measuring and inferring whether
$|\Omega_{x}\rangle=|\Omega_{y}\rangle$. The original proposals
\cite{Buh01,Bea04} provided Roger with a controlled swap gate and
an ancilla qubit (Fig.~1 (a)). The ancilla is prepared as
$(|0\rangle+|1\rangle)/\sqrt{2}$ and entangled with the
fingerprint states as follows: the two fingerprint states are not
swapped if the ancilla is in the state $|0\rangle$ and swapped
otherwise. The ancilla then passes through a Hadamard gate and is
measured in the logical basis with outcome $r\in\{ 0,1 \}$
corresponding to the ancilla being in state $|r\rangle$. This
strategy yields a one-sided error protocol because Roger's error
rate when Sapna sends $x=y$ is $p^{\text{same}}_{\text{err}}\equiv
1-\frac{1}{2}[1+\delta]=0$.
In the WCS, Sapna always sends different states so that, when Roger
obtains $r=0$, he infers $z=1$ with error rate
$p^{\text{diff}}_{\text{err}}=p^{\text{WCS}}_{\text{err}}
\equiv p_{\text{err}} =1-\frac{1}{2}[1-\delta] $.
For $m=4$ and tetrahedral encoding, we obtain
$p^{\text{WCS}}_{\text{err}}=\frac{2}{3}$ .

A controlled swap gate is not available in a linear optical
system, but we show that it is not required. If Alice and Bob each
send a single photonic qubit encoded in polarization to Roger,
then Roger only needs to measure whether the photons are in the
same polarization. This measurement can be
accomplished with the use of a Bell state discriminator that can
distinguish between $|\Psi^-\rangle$ and the other three Bell
states. Optically this discrimination is achieved by directing
each of Alice's and Bob's photons into separate input ports of a
symmetric beam splitter and observing photon count events from
two photodetectors placed at the output ports. 
A coincidence detection implies the state of the photon pair before
the beam splitter was not orthogonal to $|\Psi^-\rangle$ because 
the other three Bell states result in two photons leaving the beam 
splitter through the same port~\cite{Mat96}. 
These states exhibit a Hong-Ou-Mandel (HOM) dip
in the coincidence rate~\cite{Hon87} 
as the delay of the incidence photons is varied.

\begin{figure}
\includegraphics[scale=.6]{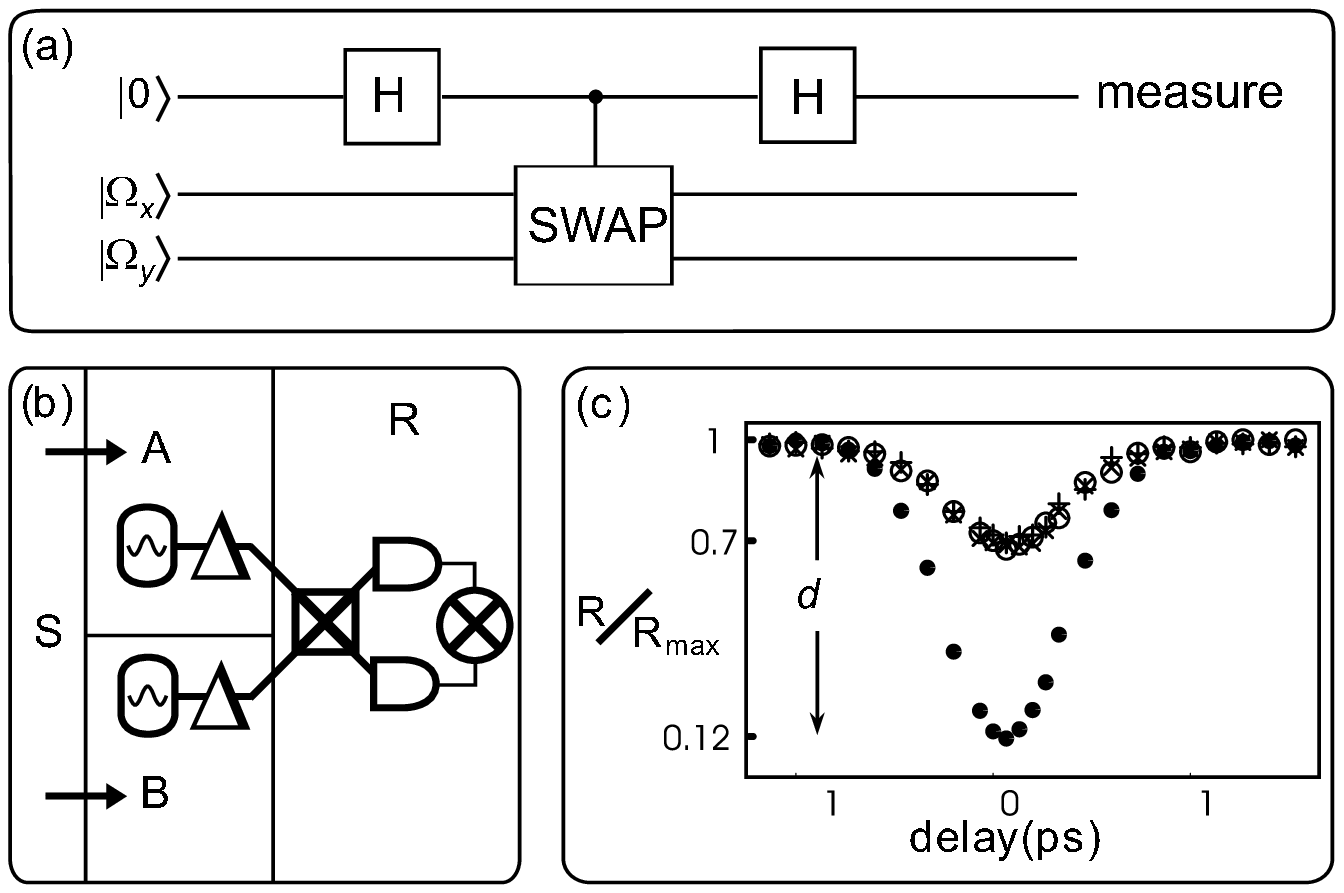}
\caption{\label{superposition-i} 
(a) Quantum circuit of the original fingerprinting protocol
\cite{Buh01,Bea04}. 
(b) Linear optical implementation: Alice (A) and Bob (B) each
receive a two-bit message from Sapna (S) and a single photon in a
known polarization state from a source. The photons are
transformed (represented by $\triangle$) to particular tetrahedral
states according to the received message. The photons are sent to
Roger (R) who mixes them at a symmetric beam splitter $\boxtimes$ and
uses coincidence detection (two detectors
$\sqsubset$\hspace{-.08in}$\supset$ and a multiplier $\otimes$) to
infer if the messages were the same or different. 
(c) Coincidence
dip with state~$|\Omega_1\rangle$ mixed with $|\Omega_0\rangle$
($\circ$), $|\Omega_1\rangle$ ($\bullet$), $|\Omega_2\rangle$ (+), and
$|\Omega_3\rangle$ ($\times$). The plots correspond to the
normalized coincidence rate $R/R_\text{max}$ 
(with $R_\text{max} = 474$ s$^{-1}$)
vs the relative delay
between two photons. The dip depth for indistinguishable photons
is labelled $d$.}
\end{figure}

For $m=4$, Alice and Bob each receive two-bit messages from Sapna,
which are used to encode their photonic qubit into one of the
tetrahedral states. Their photons are transmitted to Roger who
infers using a symmetric beam splitter whether the messages were the
same or different. This protocol is depicted in Fig.~1(b). Ideally
Alice and Bob would have separate single-photon-on-demand sources,
but practically they will be supplied with correlated, unentangled photons 
from a down-conversion source. Later we consider the case that Alice and Bob
share entangled photons.

Following the same notation as for the controlled swap case, Roger
assigns $r:=0$ for a no-coincidence and $r:=1$ for a coincidence event,
then employs (as before) the pure strategy $z=1-r$. The result
$r=1$ guarantees the messages are unequal but $r=0$ only indicates
that the messages were possibly the same. In fact this HOM dip
protocol is equivalent to the controlled swap version of single
qubit quantum fingerprinting because if Sapna sends
$x\neq y$, the probability that Alice's and Bob's photons do not
trigger a coincidence detection
is identically $p^\text{diff}_\text{err}=p_\text{err}=\tfrac{1+\delta}{2}$. 
Thus Sapna always sends different messages in the WCS.
For $m=4$, $p^\text{WCS}_\text{err}=\tfrac{2}{3}$. 

This error rate appears relatively high, yet it is superior to
classical one-bit fingerprinting with one-sided error, in which failure is
guaranteed for at least one pair of messages, resulting in a 100\% WCS 
error rate \cite{Bea04}.
Of course 100\% failure rate
for the classical case can be improved by allowing Roger a random
strategy, but then the quantum protocol can be improved in the same way,
always maintaining its superiority over the classical case~\cite{Bea04}.

\emph{Experiment.--} The feasibility of this protocol has been
illustrated by creating simultaneous pairs of
tetrahedral states and analyzing the dip achieved by Roger's
set-up in Fig.1(b).  To create correlated photons, a Ti:Sapphire
laser tuned to a wavelength of 790 nm emitted 170 fs pulses that
were frequency doubled and then down-converted in a type I
configuration via a 2-mm beta-barium borate crystal. Output
photons were spectrally filtered with a 2-nm interference filter
and transmitted through $\lambda/2$ and $\lambda/4$ waveplates
which were rotated to convert the polarization state in each
channel into one of the tetrahedral states. The two photons
were then overlapped in free space on a symmetric beam
splitter and subjected to measurements with single-photon counting
modules; the experimental results are presented in Fig.~1(c) where
state~$|\Omega_1\rangle$ is mixed with itself and each of the
other three states. The largest dip in Fig.~1(c) corresponds to
the traditional HOM dip with two identical states mixing at the
beam splitter, and the degree of distinguishability is varied by
controlling the relative delay between the two photons. The
experimental coincidence rates as a fraction of the maximum
coincidence rate~$R/R_\text{max}$ for all 16 possible fingerprint pairs
is given in Table~\ref{tab:table2} and is consistent with Clarke
{\em et al.}'s experimental results for tetrahedral states \cite{Cla01}.

\begin{table}
\caption{\label{tab:table2}Experimental visibilities of the
Hong-Ou-Mandel dip for each pair of tetrahedral states}
\begin{ruledtabular}
\begin{tabular}{cccccc}
&\vline&&Bob\\
\hline
Alice&\vline&0&1&2&3\\
\hline
0&\vline& 0.88 & 0.31 & 0.24 &0.26\\
1&\vline& 0.30 & 0.88 & 0.25 &0.40\\
2&\vline& 0.44 & 0.30 & 0.89 &0.25\\
3&\vline& 0.20 & 0.30 & 0.35 &0.89\\
\end{tabular}
\end{ruledtabular}
\end{table}

Due to birefringence in the beam splitter and limitations on
constructing the unitary transformation required to create perfect
tetrahedral states, the visibilities vary, but the dip depth $d$
for mixing identical states is consistently at 88\% or higher, as
shown in the diagonal elements of Table~1. Ideally Table~1 would
have unity for all diagonal elements (ie. $d=1$) and $1/3$ for all
off-diagonal elements. We use Bob's state~$|\Omega_1\rangle$ as
the reference state for assessing feasibility, which consistently
produces visibilities of approximately 30\% when mixed with
Alice's other three states. Note that in the non-ideal case, $d<
1$, and the error probabilities change. That is,
$p^\text{diff}_\text{err}=\frac{1}{2}(1+d\, \delta_\text{diff})$ and
$p^\text{same}_\text{err}=1-\frac{1}{2}(1+d\, \delta_\text{same})$
\cite{Hor04}, where for the tetrahedral states 
$\delta_\text{same}=1$ and $\delta_\text{diff}=1/3$.

This experimental scheme is not directly applicable to fingerprinting
because Alice and Bob are not aware when a photon pair has been produced,
so the amount of information the parties send to Roger cannot be
traced. This problem can be resolved by using either deterministic
single-photon sources or heralded single-photon sources based on two
down-converters. 
Another issue is that a large fraction of information is lost due to poor
single-photon detection efficiency. This can be overcome by Roger using
number-resolving detectors \cite{Tak99}, post-selecting 
the data on the registration of two
photons, and requesting Alice and Bob to repeat their messages if 
photons were lost.

\emph{Two-sided errors.--} The one-sided error protocol is
predicated on unitary diagonal elements of
Table~\ref{tab:table2}; as this is impossible, an experimental
protocol must be assessed for two-sided errors because, even if
Sapna sends the same messages to Alice and Bob, Roger is no longer
guaranteed to obtain the measurement outcome $r=1$. Allowing Roger
to err on both inferences lowers the classical WCS error probability bound
from 1 to $p^\text{WCS}_\text{err}\ge 0.5$ for one-bit
fingerprinting \cite{Bea04}. Thus, quantum fingerprinting is
advantageous provided that Roger's strategy yields
$p^\text{WCS}_\text{err}<0.5$ when permitting a two-sided error
protocol.

Whereas Roger followed a pure strategy for protocols with
one-sided error this restriction is unnecessary for a two-sided
error protocol. As such we introduce a procedure for producing a
successful two-sided error protocol where Roger incorporates
randomness and follows a mixed strategy instead.

The mixed strategy is as follows. Roger makes an initial inference
$z^*=1-r$ as before. If $z^*=1$ Roger infers $z=0$ with
probability $\pi_0$ and if $z^*=0$ Roger infers $z=1$ with
probability $\pi_1$.

The success rate is
\begin{align}
\pi_0&p_\text{err}^\text{diff}+(1-\pi_1)(1-p_\text{err}^\text{diff}), \nonumber \\
\pi_1&p^\text{same}_\text{err}+(1-\pi_0)(1-p^\text{same}_\text{err}),
\label{success}
\end{align}
for Sapna supplying $x\neq y$ and $x=y$ respectively. Roger then
chooses values of $\pi_0$ and $\pi_1$ such that his success rate
is identical for both cases making Sapna's choice of messages
irrelevant: all cases correspond to a WCS. We solve both success
equations~(\ref{success}) based on the values in
Table~\ref{tab:table2} where $d=0.88$,
$p^\text{same}_\text{err}\sim 0.06$, and
$p_\text{err}^\text{diff}\sim 0.65$ and find that the success rate
can achieve $0.59$ which is above the classical threshold of 0.5. 
This optimal case is achieved by
setting $\pi_0=0.37$ and $\pi_1=0$, which means that Roger's best
strategy is to treat the protocol as if it were one-sided, thereby
invoking randomness only on the side with error.

\emph{Shared entanglement.--} Thus far Alice and Bob have been
denied any communication, but experimentally it is straightforward
to provide Alice and Bob with an entangled pair of
photons. We show that shared entanglement allows perfect
single-qubit quantum fingerprinting for $m=4$ and, furthermore,
exceeds the classical limit. The classical analog to this case
corresponds to the performance in the WCS for Alice and Bob
sharing random bits that are secret from Sapna.

We allow Alice and Bob to share the Bell singlet state
$|\Psi^-\rangle$. Alice and Bob each receive a two-bit message
from Sapna and apply one of the four Pauli operations according to
which message has been sent. The result is that the state sent to
Roger is one of the four Bell states. If
Alice and Bob perform the same Pauli operation,
$|\Psi^-\rangle$ is invariant (up to a global phase); if Alice
and Bob apply different transformations, $|\Psi^-\rangle$ maps
to a different Bell state. Thus, for Roger to infer whether the
messages are the same or different, he needs only to detect
whether he has received the state $|\Psi^-\rangle$ or not. The
Bell state discriminator, in the form of a HOM dip apparatus
discussed earlier suffices as a discriminator between the Bell
state $|\Psi^-\rangle$ and the other three Bell states
\cite{Mat96}. For a perfectly efficient setup, a coincidence is
guaranteed for an input Bell state $|\Psi^-\rangle$, and no
coincidence occurs for the other Bell states. Therefore, the
protocol can achieve $p^\text{WCS}_\text{err}=0$ 
and by consuming one ebit for each pair of two-bit
messages delivered Sapna.

The physics underlying this fingerprinting scheme resembles that employed in
quantum dense coding \cite{Mat96}, but the purposes that 
these two communication protocols 
serve are quite different. Whereas, in the latter case, a
shared ebit is used to communicate a classical two-bit message from Alice to
Bob, the former allows a third party (Roger) to compare two two-bit messages.

A 100\% success rate is unachievable in classical one-bit
fingerprinting regardless of how many random bits Alice and Bob
share. If Alice and Bob share one random bit (in the case of a
shared ebit, Alice and Bob could convert their ebit to a shared
random bit if they wish), Roger's success rate for classical
one-bit fingerprinting rises from zero to $\tfrac{1}{2}$ when
Roger follows a pure strategy. If Alice and Bob share an
arbitrarily large number of random bits, Roger's success rate
improves but cannot exceed $\tfrac{2}{3}$ for any fixed number of
random bits \cite{Hor04}. Of course limited detector efficiency
for the entangled protocol will diminish the success rate, but any
success rate beyond $\tfrac{2}{3}$ is superior to the classical
case.

\emph{Conclusions.--} We have proposed an optical protocol for
single-qubit fingerprinting, experimentally demonstrated
its functionality for the case $m=4$, and shown that tetrahedral
states can be produced that meet the requirements for beating the
classical one-bit fingerprinting protocol for $m=4$. We have also
proven that single-qubit quantum fingerprinting with shared
entanglement can succeed with a zero error rate, which beats the
classical fingerprinting protocol with an arbitrary amount of
shared randomness between Alice and Bob. The experimental results
show that, in reality, two-sided errors must be accounted for, but
we have shown that Roger's best strategy is to randomly vary his
inference of whether the states are the same but not change his
guesses as to whether they are different, and this approach yields
a performance, given experimentally obtained parameters, that
exceeds the classical error bound. Quantum fingerprinting is an
excellent example of the new field of quantum communication
complexity~\cite{Bra03}, and our results here open this field to
experiments. Further work is now underway on quantum
fingerprinting with two qubits and beyond, which will allow
scaling and complexity issues to be fully investigated.

\emph{Note:--} Optical quantum fingerprinting was considered by
Massar~\cite{Mas03}, but his protocol is very different: Alice and
Bob share a single photon, and the protocol uses two-slit
interference as an alternative approach to Roger's strategy. In
our protocol Alice and Bob each have independent photons, with or
without shared entanglement. Despite the related names, the two
protocols are entirely different and with different aims: our goal
is to ensure that quantum fingerprinting operates within the
strict confines of the simultaneous message passing model,
inspired by de Beaudrap's analysis of single-qubit
fingerprinting~\cite{Bea04}.

\acknowledgments

We appreciate valuable discussions with  J.\ N.\ de Beaudrap, R.\
Cleve, and J.\ Watrous. This research has been supported by
Alberta's informatics Circle of Research Excellence (iCORE).

\end{document}